# Hole interactions with molecular vibrations on DNA

Ales Omerzu, Matjaz Licer, Tomaz Mertelj, Viktor V Kabanov, and Dragan Mihailovic

Dept. of Complex Matter, Institute Jozef Stefan, Jamova 39, 1000 Ljubljana, Slovenia

We report on a study of the interactions between holes and molecular vibrations on dry DNA using photoinduced infrared absorption spectroscopy. Laser photoexcited (PE) holes are found to have a room-temperature lifetime in excess of of $\tau > 1$ ms, clearly indicating the presence of localization. However, from a quantitative model analysis of the frequency shifts of vibrational modes caused by the PE holes, we find the hole-vibrational coupling constant to be relatively small, $\lambda \approx 0.2$. This interaction leads to a change in the conformational energy of $\Delta E_0 \sim 0.015$ eV, which is too small to cause self-trapping at room temperature. We conclude that – at least in the dry (A) form -DNA is best understood in terms of a double chain of coupled quantum dots arising from the pseudo-random chain sequence of base pairs, in which Anderson localization prevents the formation of a metallic state.

Determining the microscopic details of how electron and hole radicals interact on DNA is important not only for determining the dominant fundamental interactions on the molecule and its electronic properties as a molecular building block, but also for understanding in more detail the processes involved in DNA's life cycle. After numerous studies of the conduction properties of DNA, very diverse conclusions regarding the nature of DNA were proposed, ranging from metal to insulator and even superconductor[1,2]. Photochemical studies of electronic transport along DNA have shown that the electrons travel over short distances by quantum-mechanical tunnelling up to a range of 2-3 base pairs (i.e. up to 10 Å)[3], while transport over larger distances has been suggested to proceed via thermally activated hopping, either as polarons or by simple hopping between rigid random traps formed by the pseudorandom potential of the base-pair stacks. Recent calculations of DNA molecular dynamics have predicted that non-linear mechanics might lead to the existence of "vibrational hot spots" – selective sequences of base pairs where trapping of charge carriers can occur[4]. On the other hand, ab-initio calculations[5] suggest that geometry distortions caused by holes cause polaron formation with binding energy of the order of 0.3 eV. However, very few microscopic data on charge carrier trapping are presently available, so it is of great fundamental importance to obtain spectroscopic information on coupling of the DNA molecular vibrations with the electrons and holes. This information is crucial - not only for understanding electron and hole dynamics in photoexcited DNA - but also for understanding the mechanism of carrier transport and gene transcription in general.

Applying photoinduced infrared absorption (PIA) spectroscopy to DNA samples for the first time, we show that the PE radicals form metastable states with a unique fingerprint in the infrared spectrum. We measure infrared absorbance spectra with, and without laser illumination ($A_i$ and $A_0$ respectively), whereby the small differences in the two spectra $\Delta A = A_i - A_0$ are recorded[6]. If the lifetime $\tau$ of the photoexcited state is sufficiently long to give rise to a measurable steady-state population in the time-window of the experiment (typically $t_m \sim 1$ s), one can observe signatures in IR spectra due to the presence of *photoexcited* carriers[6]. The normalised change in absorbance $\Delta A$ (or, equivalently, the normalised differential transmittance $-\Delta T/T$) accumulated during a measurement time window $t_m$ under stationary conditions ($t_m \gg \tau$), is given by $\Delta A = G\tau$. Here $G$ is the photoexcitation rate per unit volume, determined by the incident photon flux. However, direct photo-excitation into the absorption bands of DNA by UV photons can cause indiscriminate breaking of bonds[7], and is really not suitable for a well-controlled study of intrinsic hole interactions on DNA. To avoid this problem, the present experiments were performed on dye-stained DNA. This enables charge-transfer injection of holes onto the DNA via photoexcitation on the dye molecule with visible light at 514 nm (see Fig.1). To enable cross-checking of the effect of the dye, we have used *two* different dyes: ethidium bromide (EtBr) and SYBR green (SyG). EtBr is known to intercalate in between base pairs[8], while SyG is believed to attach itself to the minor groove of the DNA[9]. For the photoinduced absorption experiments presented here, purified DNA samples of Calf Thymus (CT) and Salmon Testes (ST) DNA (Sigma) were first dissolved in deionized water. A solution of dye was added to the DNA solution. The samples were then dried to produce thin free-standing films which were then mounted into a Bomem infrared spectrometer, typically set to 4 cm$^{-1}$ resolution. Photoexcitation was performed with Ar$^+$-ion laser with $\lambda=514$ nm, using an optical fluence of $< 100$ mW/cm$^2$. The laser-on/laser-off measurement sequence had a $t_m=2$ second period.

The infrared absorbance $A_0$, and PIA $\Delta A$ of calf thymus (CT) DNA or salmon testes (ST) DNA, stained with either EtBr or SyG are shown in Fig.2. The PIA $\Delta A$ spectra are very similar to each other, but not identical. None of them bear any direct resemblance to the infrared absorbance spectrum $A_0$ however. The reproducibility of the spectra is quite remarkable, whereby the *spectral signature* does not depend significantly on dye concentration, or laser intensity. The amplitude $\Delta A$ shows clear signs of saturation as a function of dye concentration, at around 0.1 dye molecules per base pair (bp). Importantly, the CT and ST PIA spectra are virtually identical, suggesting that the photoexcited DNA has a characteristic PIA fingerprint which is independent of the DNA sequence. On the other hand, since the dyes are attached to DNA in different positions[8,9], we observe small differences in the PIA spectra in the two cases. More specifically, peaks around 1291, 1503 and 1607 cm$^{-1}$ don't appear in both dye/DNA combinations, but dye-specific and intrinsic DNA spectral signatures can be clearly and unambiguously distinguished from each other. Another possible artefact - laser heating - can be eliminated from consideration by an examination of the difference between two IR absorbance spectra obtained at different temperatures. Such a difference spectrum obtained from 295 K and 330 K spectra is shown in Fig. 2, and clearly bears no resemblance to the PIA spectra.

A surprising feature of the data – which is of great significance to mutagenic processes - is the magnitude of the PIA spectrum intensity, $\Delta A = 10^{-3}$ even at room temperature. This is a direct indication that the hole radical lifetime $\tau$ is very long. From the photon flux of $\phi = 3 \times 10^{17}$ photons/s, illuminating a 1 mg sample consisting of ~9 × 10$^{17}$ base pairs, we calculate G ≈ 0.3 photons per base pair per second. We can thus estimate the hole lifetime to be $\tau = \Delta A/G \approx 10^{-3}$ s.

We turn to a detailed discussion of the PIA spectra. First, we note the absence of a polaron hopping spectrum above 500 cm$^{-1}$, which dominates the PIA spectra of conducting polymers for example[6]. A polaronic hopping spectrum in the PIA has a peak at, or near, the polaron binding energy $E_B$. Its absence suggests that there are no polarons present with a binding energy anywhere between 0.08 eV and 0.5 eV. Instead, rather remarkably, only changes in vibrational modes are observed. The assignments of the modes are listed in Table 1. Since the PIA spectra $\Delta A$ bear no resemblance to the IR absorbance spectra $A_0$, the assignments cannot be made directly. However, when we compare the PIA spectra $\Delta A(\omega)$ in Fig. 3 with the *change* in the IR absorbance $\Delta A'(\omega, \Delta\omega) = A_0(\omega) - A_0(\omega - \Delta\omega)$, which would arise from a red-shift $\Delta\omega$ of IR modes to lower frequency, we see an obvious similarity between $\Delta A'(\omega, \Delta\omega)$ and $\Delta A(\omega)$. In Fig.3 we observe that a red-shift of $\Delta\omega = 16 \pm 2$ cm$^{-1}$ gives an impressively good overall fit to the PIA data in the region 800-1800 cm$^{-1}$. However, when individual modes are carefully analysed, we find that the shift $\Delta\omega$ is actually slightly different for different modes, as shown in Table 1.

In the region 2500-3500 cm$^{-1}$, the PIA spectrum has a large oscillator strength (approx. 60% of the total intensity). The broad peak centered at 2746 cm$^{-1}$ lies in the region of the N-H vibration band, suggesting that it is a photoinduced band, whose oscillator strength is derived from N-H vibrations, while the strong peaks at 3192 cm$^{-1}$ and 3313 cm$^{-1}$ are close in frequency to the 3240 cm$^{-1}$ and 3364 cm$^{-1}$ N-H stretching modes, implying that the PIA modes are red-shifted photoinduced N-H vibrations. The peak near 3400 cm$^{-1}$ is close to the 3421 cm$^{-1}$ O-H stretching band, implying a photoinduced enhancement of O-H vibrations, again slightly red-shifted (by 21 cm$^{-1}$) from the usual IR frequency of the O-H band. Summarising, the largest red-shift of $\Delta\omega \approx 28$ cm$^{-1}$ appears for C=O, C=C and C=N double-bond vibrations on the bases associated with the O..H and N..H hydrogen bonds at the centre of the DNA duplex, while the deoxyribose vibrations show a smaller shift of $\Delta\omega = 20$ cm$^{-1}$, and the PO$_2^-$ groups on the outer backbone region of the molecule show even smaller shifts (of $\Delta\omega = 0 \sim 6$ cm$^{-1}$), reflecting the extent of the hole wavefunction $\psi_h$. From the dependence of $\Delta A$ on EtBr doping concentration, saturation of around 0.1 EtBr/bp (Fig. 4) suggest an upper limit to an effective longitudinal radius of approximately 3-5 bps or $r_{h//} \leq 15$ Å.

From the photoinduced frequency shifts $\Delta\omega_i$ in Table 1, we can calculate the individual coupling constants $\lambda_i$ and the polaronic binding energy $E_B$. Extending Firsov's model[10], let us consider two electronic levels (corresponding to the HOMO and LUMO states of the DNA molecule) coupled linearly with infrared-active vibrational modes. We write the interaction as $H = H_0 + H_{int}$, where $H_0$ describes the potential and kinetic energies of a set of $N$ vibrational molecular oscillators with effective mass $M_i$ and frequency $\omega_i$, and the bare HOMO-LUMO electronic excitations on the $i^{th}$ site, with energy splitting $2\Delta$:

$$H_0 = \sum_{i=1}^{N}\left(\frac{1}{2}M_i\omega_i^2 x_i^2 - \frac{1}{2M_i}\frac{\partial^2}{\partial x_i^2}\right) + \Delta(a_1^+ a_1 - a_2^+ a_2) \qquad (1)$$

The interaction part $H_{int}$ between dipolar infrared-active vibrations and the electronic excitations can be written to first order as:

$$H_{int} = -\sum_i \sqrt{\lambda_i M_i \omega_i^2 \Delta}\left(a_1^+ a_2 - a_2^+ a_1\right) x_i \qquad (2)$$

where $\lambda_i$ are the coupling constants. In the classical limit, we can easily diagonalize $H$, obtaining the lowest eigenvalue:

$$E = \sum_{i=1}^{N}\frac{1}{2}M_i\omega_i^2 x_i^2 - \Delta\sqrt{1 + \left(\sum_i \alpha_i^{1/2} x_i\right)^2} \qquad (3)$$

where $\alpha_i = \lambda_i \frac{M_i \omega_i^2}{\Delta}$. For weak coupling ($\alpha < 1$), the solution to (3) for each vibrational mode individually is $x = 0$. This means that there is no shift of the equilibrium position, but only a change in vibrational frequency $\omega_i$ due to the anharmonicity in the vibrational potential. The new frequencies $\omega_i$ are given by the second derivative of $E$ with respect to $x_i$:

$$\frac{\partial^2 E}{\partial x_i^2} = M\omega_i^2 - \frac{\lambda_i M_i \omega_i^2}{\Delta^2} = M_i \omega_i^2 (1 - \lambda_i) \qquad (4)$$

The vibrational frequency is thus shifted from the unperturbed value $\omega_i$ to $\tilde{\omega}_i = \omega_i \sqrt{1-\lambda_i}$, and the normalized frequency shift $\frac{\Delta\omega_i}{\omega_i} \approx -\frac{\lambda_i}{2}$. Thus from $\Delta\omega_i/\omega_i$ we can determine the hole coupling coefficients $\lambda_i$ for each *individual* mode. Examination of Table 1 reveals that $\lambda_i$ increases towards the centre of the DNA duplex. However, the *overall* hole-vibron coupling constant is measured to be $\lambda = \Sigma_i \lambda_i = 0.21$, which is surprisingly small. From the data in Table 1, we can also calculate the renormalisation of the total ground state energy (or more concisely, the zero-point motion energy) associated with the PE radical as: $\Delta E_0 = \frac{1}{2}\sum_i \Delta\omega_i = 0.015$ eV. The value $\Delta E_0$ thus obtained is small compared to the thermal energy at 300K, indicating that radicals do not readily form self-trapped lattice polarons, contrary to some recent theoretical predictions[5].

Thus, the emerging picture of DNA, based on the present experiments, is of a complementary double chain, in which the bases form a quasi-random sequence of both longitudinally and laterally *coupled* quantum dots, where mainly the guanine (due to its significantly lower oxidation potential) acts as a hole trap (potential well). A pseudo-random base sequence gives rise to quantum wells with different confinement radii, leading to a diverse energy landscape of low-energy bound states, consistent with the existence of low-energy excitations with energies on the order of 0.1-0.3 eV [1]. Doped DNA on the other hand might be expected to form metallic segments, possibly explaining the controversially high measured conductivity under electron-beam illumination conditions[11]. We conclude by noting that the use of engineered quantum-well base sequences for storing information at room temperature might be a realistic possibility: The confinement energy $E_0 = h^2/8mL^2 \approx 0.37$ eV for a hole of radius of $r_h \approx 1$ nm, trapped in a guanine quantum well is sufficiently large to allow holes to remain trapped for times of the order of milliseconds or more – which, indeed, we experimentally observe.

**Frequencies, photoinduced shifts, assignments and vibronic coupling constants of DNA vibrations.**

| $\omega_i$ (cm$^{-1}$) | $\omega_{IR}$ (cm$^{-1}$) | $\Delta\omega_i$ (cm$^{-1}$) | $\lambda_i = 2\Delta\omega_i/\omega_i$ | Assignment [13] |
|---|---|---|---|---|
| 970 | 968 | 20 | 0.042 | Stretching vibration of deoxyribose |
| 1088, 1115 | 1088 | 0 | 0 | $PO_2^-$ symmetric stretching vibration |
| 1265 | 1240 | 6 | 0.01 | $PO_2^-$ antisymmetric stretching vibration |
| 1502 | 1493 | 6 | 0.008 | in-plane N-H bending mode |
| 1653, 1701 | 1550 - 1750 | 28 | 0.034, 0.033 | band of in-plane vibrations of C=O, C=C and C=N groups of heterocyclic bases |
| 3192, 3313 (and 2746) | 3240, 3364 (and 2766) | 50, 50 (20) | 0.031, 0.03 (0.015) | N - H stretch |
| 3400 (weak) | 3421 | 21 | 0.012 | O - H stretch |
| | | $\sum\Delta\omega_i$ = 229 cm$^{-1}$ $\approx$ 0.03 eV | $\lambda = \sum \lambda_i \approx 0.21$ | |

**Figure 1.** Optical absorption spectra of Et-Br-intercalated and non-intercalated DNA. Insert: the energy level diagram for DNA bases and EtBr dye showing the PE of a hole on EtBr dye and CT to DNA. The energies are relative to the chemical potential[12].

**Figure 2.** The IR absorbance spectrum $A_0$ of CT DNA with EtBr dye (top trace) is compared with $\Delta A$ of four different combinations of DNA and dye (CT/EtBr, ST/EtBr, CT/SyG and ST/SyG) and with the thermal difference spectrum (between 295 K and 330 K) (bottom trace). The mode assignments are given in Table 1.

**Figure 3.** $\Delta A$ in the expanded region from 800-1800 cm$^{-1}$ for CT/EtBr and CT/SyG, compared to the shifted spectrum $\Delta A'(\omega, \Delta\omega) = A_0(\omega) - A_0(\omega - \Delta\omega)$ with $\Delta\omega = 16$ cm$^{-1}$. The $\Delta\omega_i$ for individual modes are given more accurately in Table 1. (Peaks around 1291, 1503 and 1607 cm$^{-1}$ appear only in unique dye/DNA combinations, and are thus ignored.)

**Figure 4.** The amplitude $\Delta A$ as a function of EtBr concentration shows saturation behaviour.

# Fig. 1

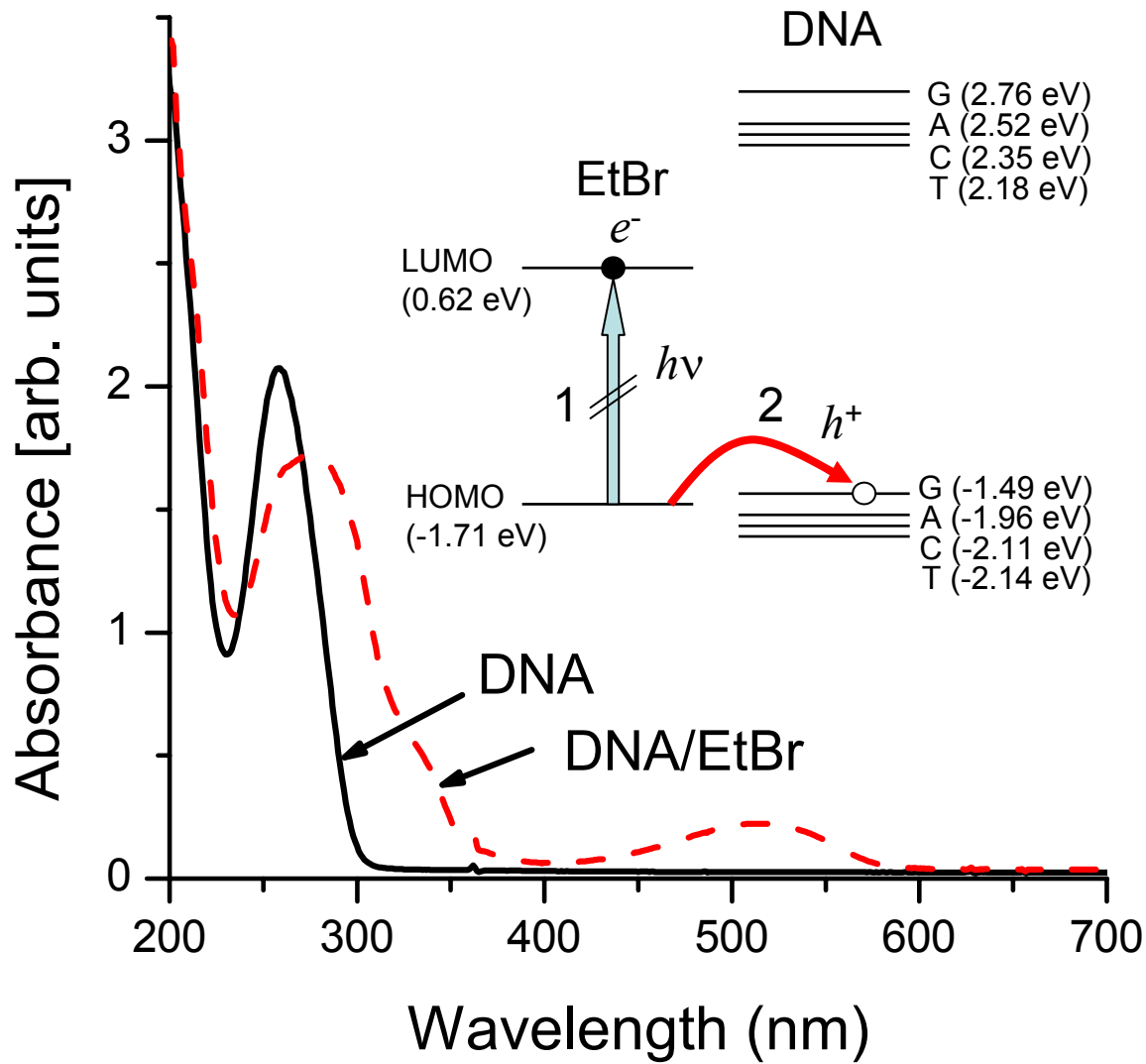

# Figure 2

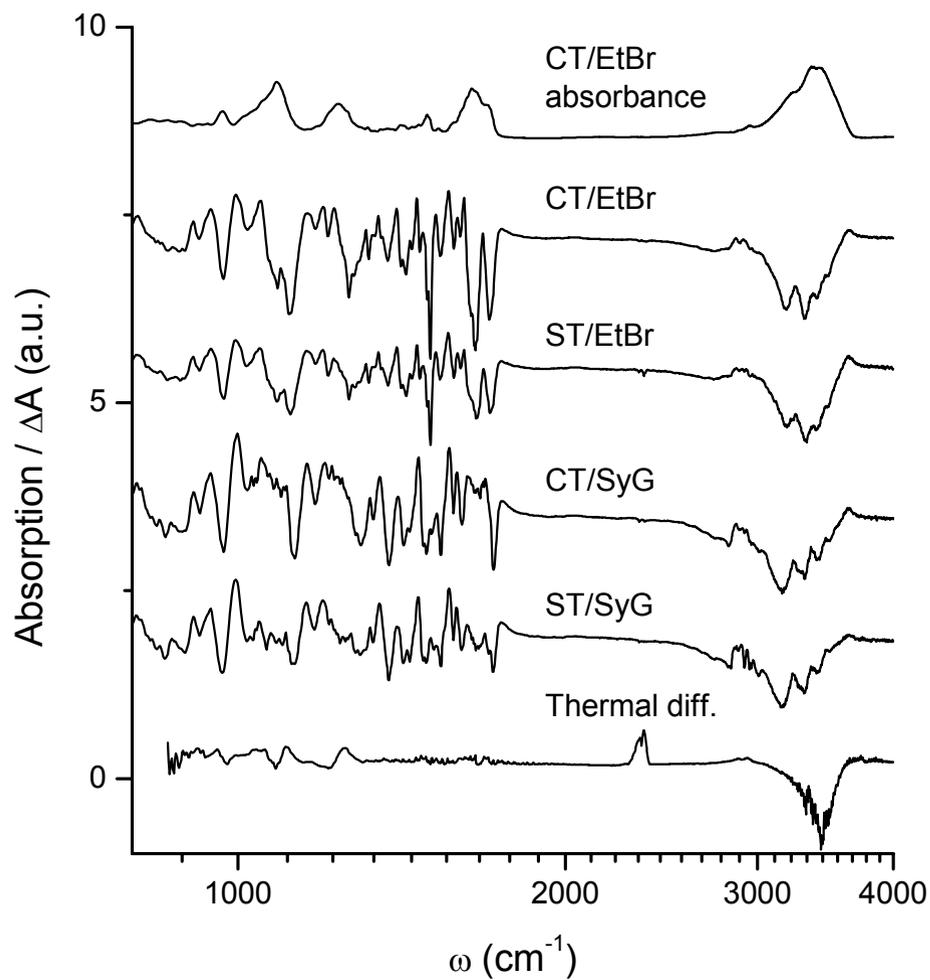

# Figure 3

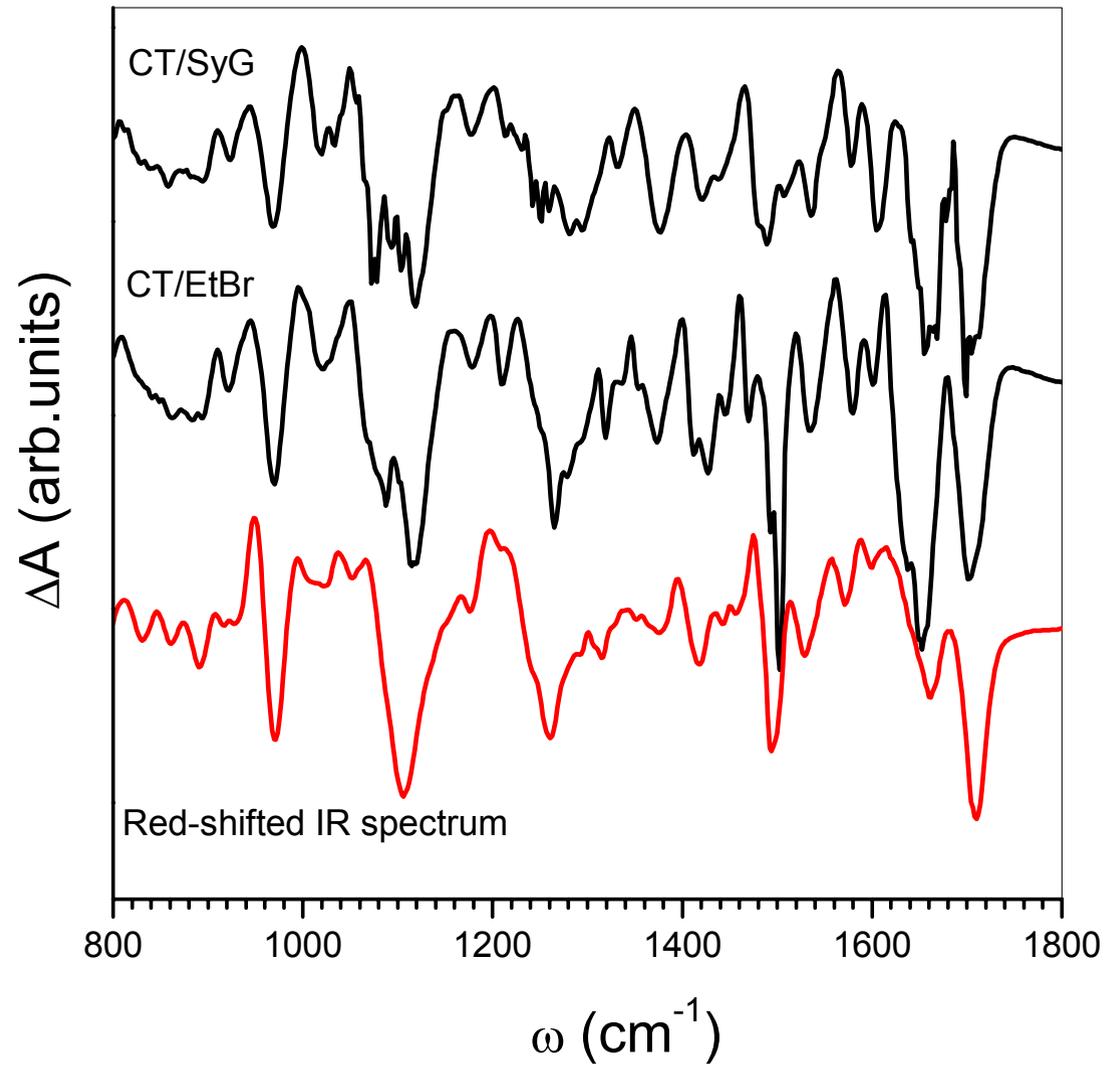

# Figure 4

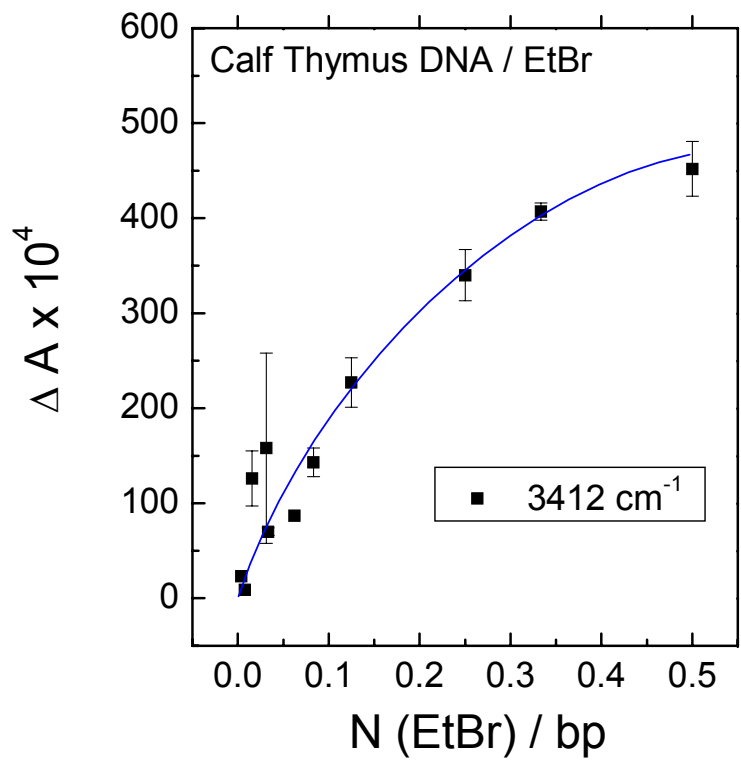